\documentclass[twocolumn,secnumarabic,amssymb,nobibnotes,aps,prl,showpacs]{revtex4-1}
\usepackage[english]{babel}
\usepackage{graphicx}
\usepackage{amsmath}
\usepackage{amsfonts}
\usepackage[usenames,dvipsnames]{color}
\usepackage{hyperref}
\hypersetup{colorlinks}
\hypersetup{linkcolor=black, citecolor=black, urlcolor=blue}
\usepackage[utf8]{inputenc}

\begin{document}
\title{Tuning colloid-interface interactions by salt partitioning}
\author{J. C. Everts}
\email{j.c.everts@uu.nl}
\author{S. Samin}
\author{R. van Roij}
\affiliation{Institute for Theoretical Physics, Center for Extreme Matter and Emergent Phenomena,  Utrecht University, Princetonplein 5, 3584 CC Utrecht, The Netherlands}
\date{\today}

\begin{abstract}
We show that the interaction of an oil-dispersed colloidal particle with an oil-water interface is highly tunable from attractive to repulsive, either by varying the sign of the colloidal charge via charge regulation, or by varying the difference in hydrophilicity between the dissolved cations and anions. In addition, we investigate the yet unexplored interplay between the self-regulated colloidal surface charge distribution with the planar double layer across the oil-water interface and the spherical one around the colloid. Our findings explain recent experiments and have direct relevance for tunable Pickering emulsions. 
\end{abstract}

\maketitle

Colloidal particles experience a deep potential well when they 
intersect fluid-fluid interfaces. They therefore adsorb strongly to such 
interfaces, self-assembling into structures such as two-dimensional monolayers 
\cite{Pieranski:1980} or particle-laden droplets in Pickering emulsions 
\cite{Pickering:1907, Leunissen:2007}. The free energy gain caused by the 
reduction of the fluid-fluid surface area is $\gamma \pi a^2 
(1+\cos\theta)^2\simeq 10^3-10^7 k_BT$, where $\gamma$ is the fluid-fluid 
surface tension, the particle radius $a$ is typically between $10-10^3$ nm, 
$\theta$ is the three-phase contact angle and $k_BT$ is the thermal energy 
\cite{Pieranski:1980}. With the exception of 
nanoparticles \cite{luo2012,reincke2006}, the binding is essentially 
irreversible and hardly prone to physicochemical modifications such as the pH or 
salt concentration. Only strong mechanical agitation is able to detach 
micron-sized particles from the interface \cite{poulichet2015}. However, the 
recovery of particles from fluid-fluid interfaces is an essential step for the 
realization of applications, such as in biofuel upgrade \cite{crossley2010}, 
``dry water'' catalysis \cite{carter2010a} and gas storage 
\cite{carter2010}.

An alternative route to overcome the difficulties 
associated with strong adsorption was offered in Refs.
\cite{Leunissen:2007, Leunissen2:2007}, which show that charged
poly(methylmethacrylate) (PMMA) particles that stabilize
water-cyclochexylbromide (CHB) Pickering emulsions are (almost) non-wetting 
($\cos\theta\rightarrow-1$), such that the colloidal particles reside 
essentially in 
the oil phase. The crucial ingredient of this system is the relatively 
``polar'' 
oil which solvates a small but significant amount of charge that stabilizes the 
colloids \cite{Leunissen:2007, Kelleher:2015}. Within a modified 
Poisson-Boltzmann theory, qualitative agreement was found with 
the experimental out-of-plane structure of the particles, provided a 
small degree of wetting was assumed \cite{Zwanikken:2007, 
Leunissen2:2007}. 

\begin{figure}[t]
\includegraphics[width=0.5\textwidth]{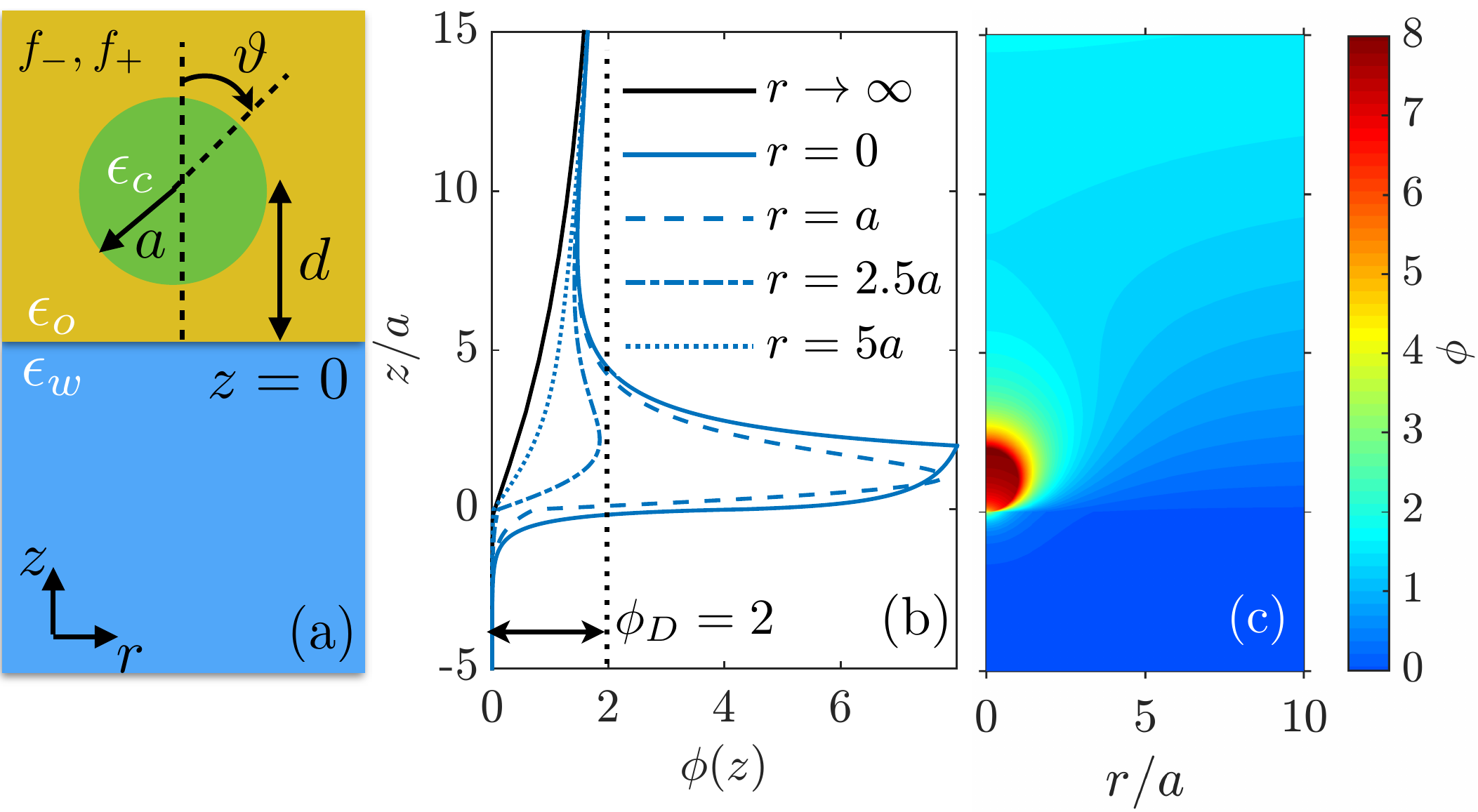}
\caption{(a) Geometry of a colloidal sphere with radius $a=1 \ \mu m$ and a 
dielectric constant $\epsilon_c=2.6$ in cylindrical coordinates 
$(r,z)$ at a distance $d$ from an oil-water interface, with the position on the 
particle surface indicated by the polar angle $\vartheta$. The oil 
and water are characterized by dielectric constants $\epsilon_o=7.92$ and 
$\epsilon_w=80$, respectively. The self-energies $(f_+,f_-)=(6,10)$ in units of 
$k_BT$ determine the degree of ion partitioning among the two solvents. In (b) 
we show the electrostatic potential $\phi({r,z})/\beta e$ for a positively 
charged sphere at $d=a$ along the $z$ direction at fixed $r$, with the Donnan 
potential $\phi_D$ indicated, and as a surface plot in (c). We set the 
charge-regulating equilibrium constant $K_+a^3=1$ ($pK_+=8.8$) and the 
screening 
length in oil $\kappa_o^{-1}=10\ \mu\mathrm{m}$.}
\label{fig:geomex}
\end{figure}

Very recent experiments on the same 
system  by Elbers \textit{et al.} \cite{Elbers:2016} 
 revealed that the colloidal particles in fact do \emph{not} penetrate the 
water-CHB interface, but are trapped at a finite $\sim$ nanometer distance \cite{Elbers:2016, suppA}, 
completely circumventing the irreversible wetting effects described above. 
Additionally, these non-touching particles were easily detached from the 
oil-water interface by the addition of an organic salt \cite{suppB, Elbers:2016}, while at the same time 
reversing the sign of the colloidal charge, see Ref. \cite{Elbers:2016}. Interestingly, this tunability offers an appealing route for controlled destabilization of Pickering emulsions. The authors of  \cite{Elbers:2016} hypothesized that the trapping mechanism is due to a force balance between 
an attractive image-charge and a repulsive van der Waals (VdW) force, as proposed by Oettel \cite{Oettel:2007}. While correctly predicting the non-touching behaviour, this argument cannot explain how the particle-interface interaction can be tuned from attractive to long-ranged repulsive by adding salt.  

The findings of Ref. \cite{Elbers:2016} show that whether a 
particle ever {\em arrives} at the interface is subtle and tunable. In this 
Letter, we show that the interaction between a charged colloidal particle and 
the 
oil-water interface consists not only of well-known image-charge \cite{Jackson, 
Wang:2012} and dispersion forces \cite{Oettel:2007, parsegian}, but also of 
particle-ion forces that can be tuned between strongly repulsive 
and strongly attractive by (i) varying the \emph{sign} of the 
particle charge and (ii) by varying the {\em difference} between the degree of 
hydrophilicity of the cations and anions. This difference determines not only 
the distribution of ions among the water and oil phases (``ion 
partitioning"), but also the sign of the Donnan potential that spontaneously 
forms between the water and oil bulk phases (not unlike the potential at a PN 
junction) 
\cite{Leunissen2:2007,Levin:2000, Bier:2008, Onuki:2008, 
Zwanikken:2008,Westbroek:2015}. In fact, our calculations show contrary to a 
common assumption \cite{Danov:2006, Frydel:2007, Oettel:2007, Wirth:2014}
that the low but non-zero ion concentrations in oil are crucial and should not be neglected \cite{Netz:1999, wurger:2006,Morrison:1993, 
Royall:2006, Boon:2015, Smith:2015, Linden:2015}, even if the ions initially present in the oil strongly prefer to be in the water phase \cite{Levin:2000, Bier:2008, 
Onuki:2008, Zwanikken:2008, bier:2011, bier:2012, onuki:2006, Onuki:2011, 
Samin:2011, Levin:2012, Samin:2013, Samin:2014, Westbroek:2015}.

To describe the coupling between the particle, oil-water interface, and ions, we will focus for simplicity on a single oil-dispersed charged 
colloidal sphere of radius $a=1\ \mu\mathrm{m}$, charge $Ze$, and dielectric constant 
$\epsilon_c=2.6$ (PMMA) with its center at a distance $d$ from a planar 
oil-water interface \cite{note1} that separates the oil (CHB) phase ($z>0$, 
dielectric constant 
$\epsilon_o=7.92$) from the water phase ($z<0$, dielectric constant 
$\epsilon_w=80$), as sketched in Fig. 1(a). Since 
$\epsilon_c<\epsilon_o<\epsilon_w$, the VdW interaction between 
particle and the oil-water interface is repulsive 
\cite{Oettel:2007, parsegian} for $d>a$. This repulsion is significant only for 
$d-a\lesssim10\ \mathrm{nm}$ given the Hamaker constant of $\sim-0.3\ k_BT$ 
\cite{Elbers:2016}. However, the repulsion is sufficiently strong to impede 
adsorption to the interface \cite{Oettel:2007}, see Ref. \cite{supp1}. Since $\epsilon_o<\epsilon_w$ the image 
charge potential 
$\Phi_\text{im}(d)=-\beta^{-1}Z^2\lambda_B^o(\epsilon_w-\epsilon_0)/[
4d(\epsilon_w+\epsilon_0)]$ (where $\beta^{-1}=k_BT$), which holds in the 
absence of salt for $\epsilon_c=\epsilon_o$ but which was shown to be accurate 
within a few percent also for non-index-matched colloidal 
particles \cite{Danov:2006}, is attractive regardless the sign of $Z$ for oil-dispersed colloidal particles, but is repulsive when the particles are dispersed in the water phase \cite{Wang:2012}. Here, 
$\lambda_B^o=\beta e^2/4\pi\epsilon_{\text{vac}}\epsilon_o$ is the Bjerrum 
length in oil. A force balance between 
the image and VdW force shows that it is possible to trap the 
particle at a finite, but small, distance from the interface 
\cite{Oettel:2007, Elbers:2016}, circumventing the non-tunable 
nature of the wetting effects. Incorporating the VdW force in 
our theory (unlike Ref. \cite{Leunissen2:2007}), we show how screening and ion 
partitioning makes the trapping mechanism highly tunable.

We use the framework of classical density functional theory to construct the grand potential 
functional $\Omega[\rho_\pm,\sigma; d]$, with $\rho_\pm({\bf r})$ the density 
profiles of (non-surface-bound) monovalent ions and $\sigma({\bf r})$ the 
particle surface charge density. Denoting the region outside the 
particle by $\mathcal{R}$ and its surface by $\Gamma$, $\Omega$ is 
given by

\begin{align}
&\beta\Omega=\sum_\alpha\int_\mathcal{R}d^3{\bf r} \ \rho_\alpha({\bf r})\left[\ln\left(\frac{\rho_\alpha({\bf r})}{\rho_s^w}\right)-1+\beta V_\alpha(z)\right] \label{eq:dft} \\
&+\int_\Gamma d^2{\bf r} \ \bigg{(}\!\!\pm\sigma({\bf r})\left\{\ln [\pm\sigma({\bf r})a^2]+\ln\left( \frac{K_\pm}{\rho_s^w}\right)+\beta V_\pm(z)\right\} \nonumber \\
&+\left[\sigma_m \!\mp\!\sigma({\bf r})\right]\ln\left\{[\sigma_m\mp\sigma({\bf r})]a^2\right\}\!\!\bigg{)} +\frac{1}{2}\int_\mathcal{R}\!d^3{\bf r}\ \!Q({\bf r})\phi({\bf r}).
\nonumber
\end{align}
The first term is the ideal gas grand potential of the ions coupled 
to the external potential $V_\alpha(z)=\beta^{-1}f_\alpha\Theta(z)$ 
($\alpha=\pm$), with $\Theta(z)=[1+\tanh(z/2\xi)]/2$, $\xi\sim 10^{-3}a$ the 
interface thickness and $\rho_s^w$ the bulk density in water. The preference of 
ions for water or oil is modeled by the energy cost $\beta^{-1}f_\alpha$ to 
transfer a single ion from the water to the oil phase. To mimic that cations are 
typically less hydrophilic than anions, we set $(f_+,f_-)=(6,10)$. The second 
term describes the free energy of a two-dimensional binary lattice 
gas of neutral and (either positively \emph{or} negatively) charged groups, with 
a maximum charge density $\sigma_m a^2=10^6$ (one charged group per nm$^2$). The 
non-electrostatic free energy of binding an ion is characterized by $k_BT\ln 
(K_\pm/ 1 \ \mathrm{M})$, with equilibrium constant 
$K_\pm=[\text{S}][\text{X}^{\pm}]/[\text{SX}^{\pm}]$ (or 
$pK_\pm=-\log_{10}(K_\pm/1\ \text{M})$) that describes the adsorption of a 
negative \emph{or} positive ion $\text{X}^\pm$ to a neutral surface site S, i.e. 
$\text{S}+\text{X}^{\pm}\leftrightarrows \text{SX}^{\pm}$. The electrostatic 
energy is described within mean-field theory in the last term, with the total charge 
density $Q({\bf r})=\rho_+({\bf r})-\rho_-({\bf r})+\sigma({\bf r})\delta(|{\bf 
r}-d{\bf e}_z|-a)$, and the electrostatic potential $\phi({\bf r})/\beta e=25.6 
\ \phi({\bf r}) \ \text{mV}$.

From the Euler-Lagrange equations $\delta\Omega/\delta\rho_\pm({\bf r})=0$ we find the equilibrium profiles $\rho_\pm({\bf r})=\rho_s(z)\exp[\mp\phi({\bf r})\pm\Theta(z)\phi_D]$, where $\rho_s(z)=\rho_s^w$ for $z<0$ and $\rho_s(z)=\rho_s^o$ for $z>0$, where $\rho_s^o=\rho_s^w\exp[-( f_++f_-)/2]$ the bulk ion density in oil. We defined the Donnan potential $\phi_D/\beta e$, with $\phi_D=(f_--f_+)/2$, which is the potential difference between the bulk oil and water phases due to ion partitioning. Combining our expressions for $\rho_\pm({\bf r})$ with the Poisson equation for the electrostatic potential, we obtain the Poisson-Boltzmann equation for ${\bf r}\in\mathcal{R}$,
\begin{equation}
\nabla\cdot[\epsilon(z)\nabla\phi({\bf r})]/\epsilon_o=\kappa(z)^2\sinh[\phi({\bf r})-\Theta(z)\phi_D],
\label{eq:PB}
\end{equation}
where $\epsilon(z)=(\epsilon_o-\epsilon_w)\Theta(z)+\epsilon_w$. Furthermore, 
$\kappa(z)^2=8\pi\lambda_B^o\rho_s(z)$, with $\kappa^{-1}(z\rightarrow\infty)$ 
the Debye screening length in the bulk oil $\kappa_o^{-1}$. We fix 
$\kappa_o^{-1}=10a$, close to typical values in CHB 
\cite{Linden:2015}, from which the screening length in water follows as 
$\kappa_w^{-1}=\sqrt{\epsilon_w/\epsilon_o}\exp[-(f_++f_-)/4]\kappa_o^{-1}
=0.58a$, which is on the high side but convenient for our numerical calculation, 
(the precise value of $\kappa_w^{-1}$ is unimportant for the physics in oil 
discussed below). Inside the dielectric particle the 
Poisson equation in the absence of external charges reads 
$\nabla^2\phi=0$. On the particle surface, we have the boundary 
condition 
${\bf 
n}\cdot[\epsilon_c\nabla\phi|_\text{in}-\epsilon_o\nabla\phi|_\text{out}]
/\epsilon_o=4\pi\lambda_B^o\sigma({\bf r})$, with $\bf n$ an outward pointing 
normal vector and where $\sigma({\bf r})$ follows from 
$\delta\Omega/\delta\sigma({\bf r})=0$ \cite{Zoetekouw, Boon:2011}, resulting for ${\bf 
r}\in\Gamma$ in the Langmuir adsorption isotherm \cite{Ninham:1971}, 
$\sigma({\bf r})=\pm\sigma_m\left\{1+K_\pm\exp[\pm(\phi({\bf 
r})-\phi_D)]/\rho_s^o\right\}^{-1}.$ Eq. \eqref{eq:PB} is solved 
numerically for $\phi({\bf r})$, from which $\rho_\pm({\bf r})$ and $\sigma({\bf 
r})$ follow, and after insertion into Eq. \eqref{eq:dft}, we find the 
particle-interface interaction Hamiltonian 
$H(d):=\min_{\rho_\pm,\sigma}\Omega[\rho_\pm,\sigma; d]$.
$H(d)$ does not contain the particle-interface VdW repulsion, which 
can be added separately \cite{supp1}. 

\begin{figure}[t]
\includegraphics[width=0.5\textwidth]{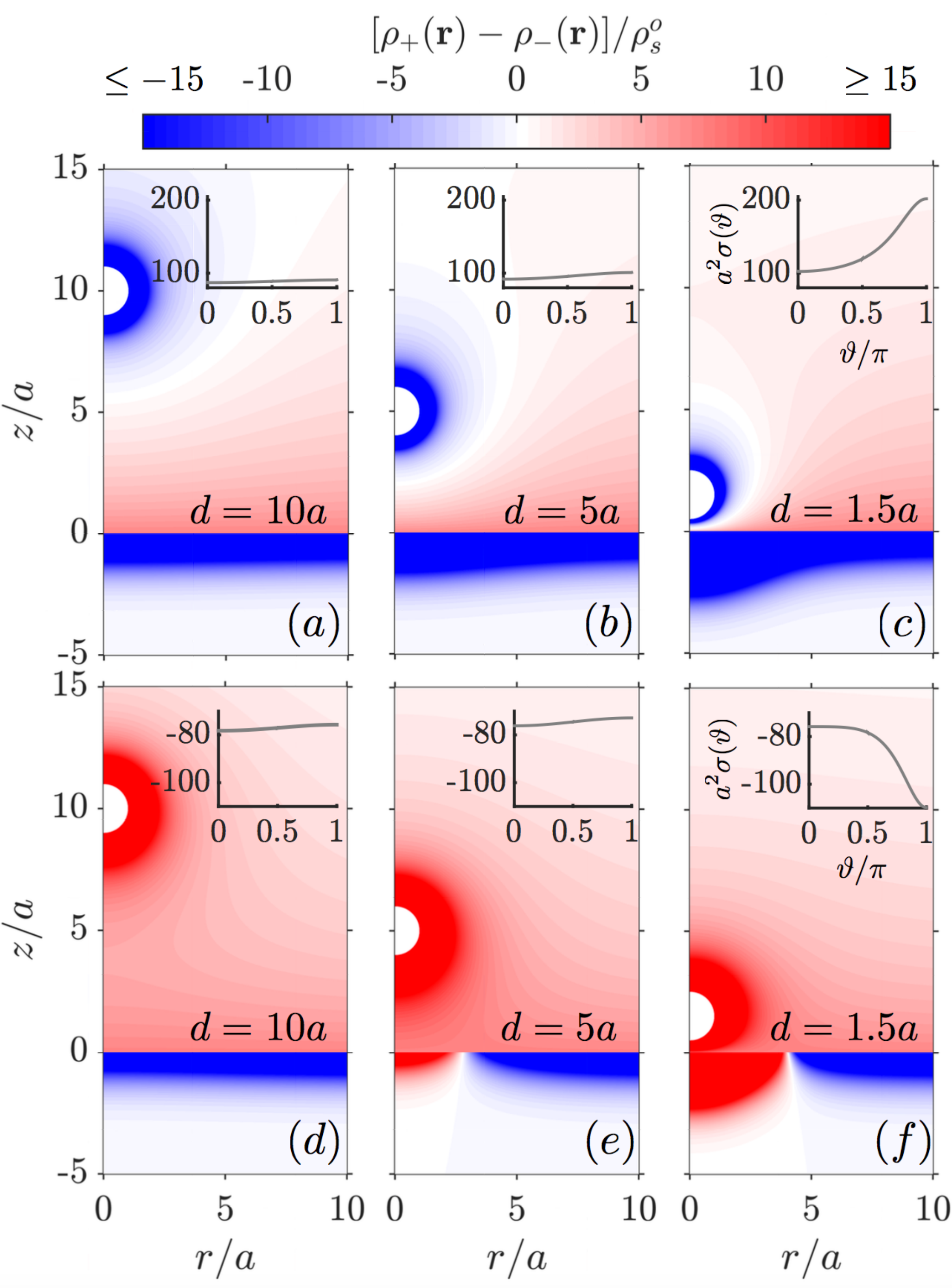}
\caption{(Color online) Double layer destruction for a charged  
particle with radius $a=1 \ \mu\text{m}$, with self-energies $(f_+,f_-)=(6,10)$, oil screening length $\kappa_o^{-1}=10a$, equilibrium constant $K_\pm a^3=1$ ($pK_\pm=8.8$), and bulk salt concentrations 
$\rho_s^0=0.093\ \text{nM}$ (oil) and $\rho_s^w=0.28\ \mu\text{M}$ (water). In (a)-(c) the particle 
is positively charged $(Z>0)$, and in (d)-(f) negatively charged $(Z<0)$.  For clarity, we introduced a 
cutoff for the scaled net ion 
density $[\rho_+({\bf r})-\rho_-({\bf r})]/\rho_s^o$. The insets 
show the particle charge distribution $a^2\sigma(\vartheta)$.}
\label{fig:destruct}
\end{figure}

An example of the resulting potential distribution around a positively 
charged colloidal sphere can be found in Fig. \ref{fig:geomex}(b)-(c), which 
reveals how $\phi({\bf r})$ approaches its asymptotic value $\phi_D$ at 
$z\rightarrow\infty$ for various axial distances $r$ from the particle, 
revealing a strong coupling between the spherical and planar geometry of the 
particle and the interface, respectively. Since $f_+<f_-$ (and 
hence $\phi_D>0$), the oil side is positively charged and the water side is 
negatively charged. This strong coupling is further illustrated in Fig. 
\ref{fig:destruct}, where the (scaled) ion density is plotted for distances 
$d=10a$, $5a$ and $1.5a$, in (a)-(c) for a positively charged 
particle, and in (d)-(f) for a negatively charged 
particle. Upon approaching the interface, the particle 
double layer deforms as ions are stripped by the water phase, since they 
dissolve better in water. As a consequence, the initially planar double layer at 
the water side strongly deforms as well. This double layer destruction was 
investigated earlier for a dense laterally averaged monolayer 
\cite{Zwanikken:2007}, but here we laterally resolve the spatial structure of 
the double layers for the first time, even with charge regulation taken into 
account. 

\begin{figure}[t]
\includegraphics[width=0.45\textwidth]{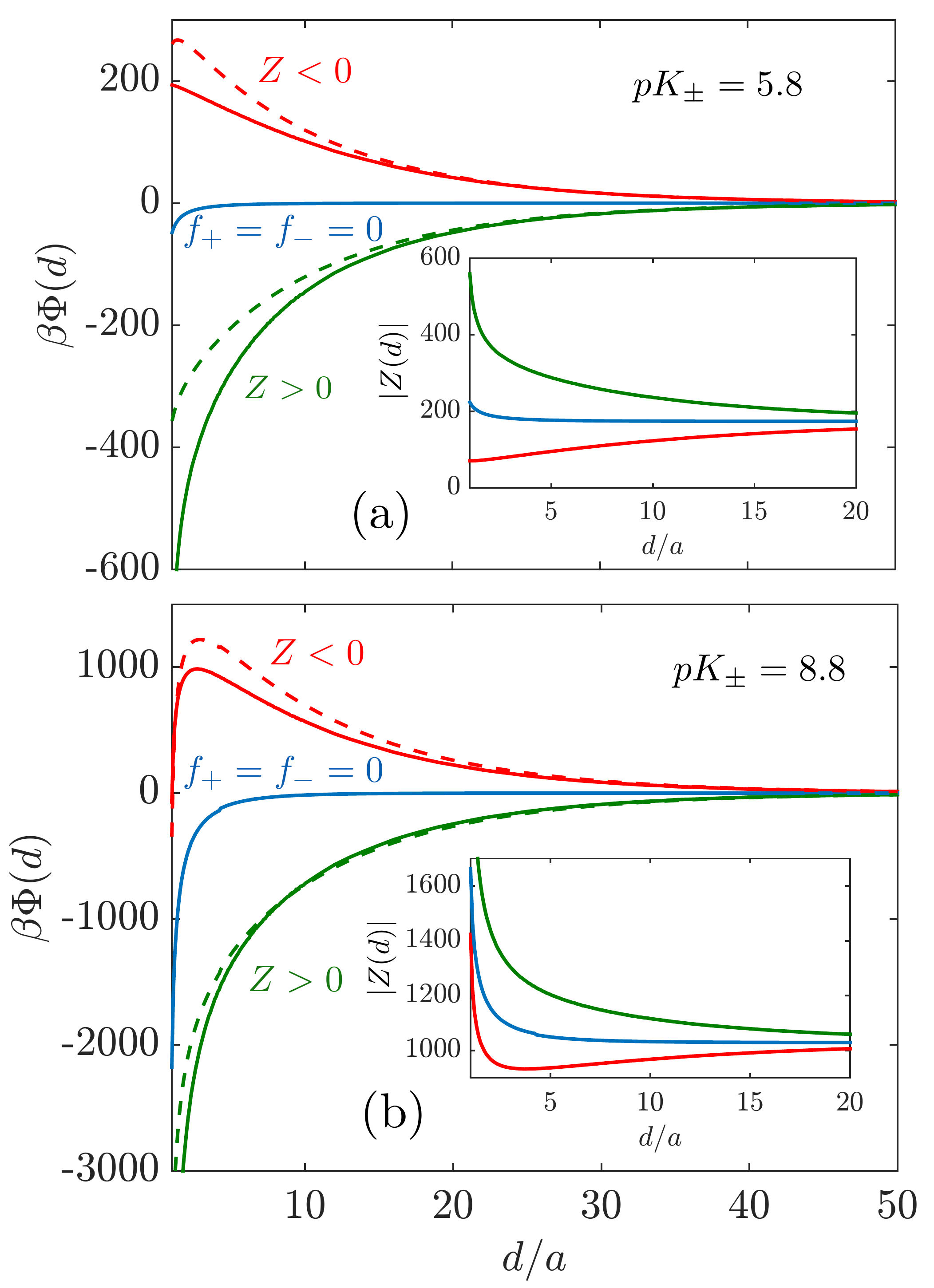}
\caption{(Color online) Particle-interface potential $\Phi(d)$ for 
oil screening length $\kappa_o^{-1}=10a$. The full curves are the results for (a) $pK_\pm=5.8$ 
and (b) $pK_\pm=8.8$. The red (green) curves are for a negative (positive) 
particle with self-energies $(f_+,f_-)=(6,10)$, the blue curves are for 
$f_+=f_-=0$ and are independent of the sign of $Z$. The insets show the 
absolute charge $|Z(d)|$. The green ($Z>0$) and red ($Z<0$) dashed curves show 
the approximation $\tilde{\Phi}(d)$. }
\label{fig:int}
\end{figure}

Interestingly, the surface charge distributions $\sigma(\vartheta)$ of negative 
and positive charge-regulating particles, shown in the insets of 
Fig. \ref{fig:destruct}, are \emph{not} related by 
$\sigma(\vartheta)\leftrightarrow -\sigma(\vartheta)$ because the antisymmetry 
is broken by $f_+\neq f_-$. Here, $\vartheta$ is the polar angle defined in Fig. 
\ref{fig:geomex}(a). For $d\gg\kappa_o^{-1}$, the particle double 
layer has spherical symmetry, and $\sigma(\vartheta)$ is constant. Close to the 
interface, however, we unravel an interplay between mass-action and 
image-charge effects, which for $f_->f_+$ enhance each other for positive 
particles, and counteract each other for negative 
particles. The enhanced cation (reduced anion) concentration close 
to the interface enhances (reduces) $|\sigma(\vartheta)|$ at the south pole 
$\vartheta=\pi$ for positive (negative) particles by mass action. 
The image-charge effect, in contrast, is independent of the charge sign and 
allows charge-regulating particles to lower their electrostatic 
energy by increasing $|Z|$ when the particle approaches a medium 
with a higher dielectric constant. Indeed, close to the interface the dielectric 
effect dominates and for negative particles the south pole is 
actually higher charged ($\vartheta=\pi$) than the north pole ($\vartheta=0$), 
see the inset in Fig. \ref{fig:destruct}(f), in contrast to the case farther from 
the interface (Fig. \ref{fig:destruct}(d),(e)). The interplay of mass action and 
dielectric effects is also seen in the total particle charge 
$Z=\int_\Gamma d^2{\bf r} \ \sigma({\bf r})$, which depends on $d$ as shown in 
the insets in Fig. \ref{fig:int} for low and high $|Z|$ in (a) and (b), 
respectively. This is of interest not only for $(f_+,f_-)=(6,10)$, but also for 
$f_+=f_-=0$ (blue curves), to isolate the image-charge effects by switching off 
ion partitioning. Mass action thus increases (decreases) $|Z(d)|$ when $Z>0$ 
$(Z<0)$, while dielectric effects always increase $|Z(d)|$. The image-charge 
effect is weak at low $|Z|$, but strong enough to drive negatively charged 
particles even more negative close to the interface. Combined with 
ion partitioning, this yields a minimum in $|Z(d)|$, if $Z$ is sufficiently negative (Fig. 
\ref{fig:int}(b)).

In Fig. \ref{fig:int} we plot the particle-interface interaction potential 
$\Phi(d)=H(d)-H(\infty)$, which is repulsive for $Z<0$ (red curve) and 
attractive for $Z>0$ (green curve) provided that $f_+< f_-$. The blue curves, 
for $f_+=f_-=0$, show a much smaller $|\Phi(d)|$; i.e., the 
particle-interface interaction is dominated by ionic rather than image-charge-
like effects. In the supplemental material \cite{supp1} we added the 
\emph{repulsive} VdW potential to $\Phi(d)$ to show that the particles can indeed be trapped a few 
nanometers from the interface. Our calculations also reveal that it is possible 
to impede adsorption in the case of an \emph{attractive} VdW interaction, since 
for $\phi_D>0$ and $Z<0$ (or as we shall see later also for $\phi_D<0$ and 
$Z>0$) there is a wetting-preventing energy barrier whose location and height 
can be tuned by the magnitude of $|Z|$ and/or $\phi_D$. 


To further analyze the particle-interface interactions, we consider 
$\tilde{\Phi}(d)= Z(\infty)\phi_0(d)+\Phi_\text{di}(d)$, with $\phi_0(z)$ the 
(analytically available) Gouy-Chapman-like electrostatic potential \emph{without} any 
particle present \cite{Westbroek:2015} as plotted in Fig. 
\ref{fig:geomex}(b) (black solid line, note that $\phi_0(z\rightarrow\infty)=\phi_D$), and 
$\Phi_\text{di}(d)=\Phi(d)|_{f_+=f_-=0}$, the purely dielectric numerically 
obtained interaction potential. In this empirical approximation for $\Phi(d)$, 
$\phi_0(d)$ acts as an external (electric) potential for a particle 
with charge $Z(\infty)$. The dashed green and red curves in Fig. 
\ref{fig:int} show that $\tilde{\Phi}(d)$ is a good approximation for $\Phi(d)$. 
Hence, $\kappa_o^{-1}$ is the relevant length scale for the 
particle-interface interaction, confirming that $\Phi(d)$ is 
insensitive to the precise value of $\kappa_w^{-1}$. It turns out that 
$\tilde{\Phi}$ approximates $\Phi$ for constant-charge particles perfectly if 
$|Z|$ is low \cite{supp2}.


To estimate whether $Z(\infty)\phi_0(d)$ or $\Phi_\text{di}(d)$ is dominant, we equate both terms in absolute value at contact, 
$|Z(\infty)\phi_0(a)|=\Phi_\text{di}(a)$, and approximate 
$\Phi_\text{di}(a)\approx\Phi_\text{im}(a)$ with $\epsilon_w\gg\epsilon_o$. The 
latter approximation gives an upper bound for the image-charge attractions, 
since salt will enlarge screening if no ion partitioning is included. We 
find that ion partitioning dominates image charges when 
$|Z(\infty)|\lambda_B^o/4a$ is small compared to $|\phi_D|$, as is the case in Fig. 
\ref{fig:int}(a). In Fig. \ref{fig:int}(b), we have 
$|Z(\infty)|\lambda_B^o/4a\sim|\phi_D|$, such that dielectric effects and ion 
partitioning are both important, although the dielectric effects are  
significant only for $d<\kappa_o^{-1}$. For $|Z(\infty)|\lambda_B^o/4a\gg|\phi_D|$, 
dielectric effects would dominate; however, it is hard to enter this regime 
experimentally, as it would require either fine-tuning the self-energies 
$f_\alpha$ to obtain extremely small Donnan potentials or particle 
charges exceeding $\mathcal{O}(10^4)$ in oil. Hence, the tunable Donnan 
potential is important in most experimental setups, and can be measured 
\cite{Vis:2014}. Moreover, we note that changing the sign of $Z(\infty)$ is 
equivalent to interchanging $f_+\leftrightarrow f_-$, changing only 
the sign of $\phi_D$, showing that the nature of the 
particle-interface interactions can also be tuned by changing the 
\emph{type} of salt. Finally, when multiple ions are included, $\phi_D$ will 
depend not only on the self-energies, but also on the bulk ion 
\emph{concentrations} \cite{Westbroek:2015}, extending even further the tunability 
options.


This Letter thus shows the importance of a low, but non-zero salt concentration in 
oil for the particle-interface interaction of an oil-dispersed 
particle. We showed that tuning the sign of the 
particle charge or Donnan potential can change attractive 
particle-interface interactions into repulsions. This sheds light 
on the very recent experimental studies by Elbers \emph{et al.} 
\cite{Elbers:2016}, who showed not only that adding a salt with $f_+<f_-$ 
to the oil changes $Z>0$ to $Z<0$, but also that particles are repelled from 
the interface to the bulk. Our calculations support this observation, since 
attractions become repulsions (Fig. \ref{fig:int}) upon changing the sign of the 
particle charge, if a suitable charge regulation mechanism is 
provided. For example, a binary adsorption model in a three-ion 
system can be used, where global charge neutrality ensures that the cation and 
anion concentrations can be varied independently. Indeed, in the experiment of 
Ref. \cite{Elbers:2016}, there are three ions, namely H$^+$, Br$^-$, and the organic 
tetrabutylammonium cation, and it is hypothesized that only H$^+$ and 
Br$^-$ can adsorb on the particle. Moreover, the range of the repulsion upon 
inversion of the particle charge is of the order $\kappa_o^{-1}$, much larger 
than the range of the VdW repulsion, that was proposed earlier as the 
responsible mechanism \cite{Oettel:2007, Elbers:2016}. Our findings also have repercussions for a VdW \emph{attraction}, since the Donnan 
potential impedes adsorption of oil-dispersed particles by a 
salt-tunable energy barrier for $Z(\infty)\phi_D<0$. Our work is relevant for designing tunable and reversible 
Pickering emulsions, which can be applied in drug delivery and food processing 
\cite{Chevalier:2013, Tang:2015}, but also for novel experiments where an ion flux can induce repulsive or attractive surface-specific interactions depending on the surface chemistry of the suspended particles \cite{Squiresa:2016}.

\begin{acknowledgments}
We acknowledge fruitful discussions with N. Elbers, J. van der Hoeven and A. van Blaaderen, and financial support of a Netherlands Organisation for Scientific Research (NWO) VICI grant funded by the Dutch Ministry of Education, Culture and Science (OCW) and from the European Union's Horizon 2020 programme under the Marie Sk\l{}odowska-Curie grant agreement No. 656327. This work is part of the D-ITP consortium, a program of the Netherlands Organisation for Scientific Research (NWO) funded by the Dutch Ministry of Education, Culture and Science (OCW).
\end{acknowledgments}

\bibliographystyle{apsrev4-1} 
\bibliography{literature1} 

\end{document}